\documentclass[aps,pra,twocolumn,superscriptaddress]{revtex4}
\usepackage{amsfonts, amssymb, amsmath,graphicx}

\begin{document}

\title{Supercurrent and dynamical instability of spin-orbit-coupled ultracold Bose gases}

\author{Tomoki Ozawa}
\affiliation{
INO-CNR BEC Center and Dipartimento di Fisica, Universit\`a di Trento, I-38123 Povo, Italy
}%

\author{Lev P. Pitaevskii}
\affiliation{
INO-CNR BEC Center and Dipartimento di Fisica, Universit\`a di Trento, I-38123 Povo, Italy
}%
\affiliation{
Kapitza Institute for Physical Problems, RAS, Kosygina 2, 119334 Moscow, Russia
}%

\author{Sandro Stringari}
\affiliation{
INO-CNR BEC Center and Dipartimento di Fisica, Universit\`a di Trento, I-38123 Povo, Italy
}%

\date{\today}

\def\del{\partial}
\def\p{\prime}
\def\simge{\mathrel{%
         \rlap{\raise 0.511ex \hbox{$>$}}{\lower 0.511ex \hbox{$\sim$}}}}
\def\simle{\mathrel{
         \rlap{\raise 0.511ex \hbox{$<$}}{\lower 0.511ex \hbox{$\sim$}}}}
\newcommand{\feynslash}[1]{{#1\kern-.5em /}}

\begin{abstract}
We investigate the stability of supercurrents in a  Bose-Einstein condensate with one-dimensional spin-orbit and Raman couplings.
The consequence of the lack of Galilean invariance is explicitly discussed.
We show that in the plane-wave phase, characterized by a uniform density, the supercurrent state can become dynamically unstable, the instability being associated with the occurrence of a complex sound velocity, in a region where the effective mass is negative.
We also discuss the emergence of energetic instability in these supercurrent states.
We argue that  both the dynamical and the energetic instabilities in these systems can be generated experimentally through excitation of the collective dipole oscillation.
\end{abstract}

\maketitle

\section{Introduction}

A characteristic feature of superfluids is their ability to  support a current flow (supercurrent) without dissipation~\cite{Kapitza1938,Allen1938}. In a Galilean invariant, uniform configuration the supercurrent  does not decay if the velocity of the fluid is lower than the critical velocity fixed by the famous Landau criterion~\cite{Landau1941,Lifshitz1980} $v_{\mathrm{cr}}= \min \epsilon(p)/p$ where $\epsilon(p)$ is the excitation energy of the elementary excitation of the system carrying momentum $p$. Landau's instability has an energetic nature, being associated with a negative value of the excitation energy. 
Nonuniform superfluid systems, like ultracold Bose-Einstein condensates in optical lattices, are known also to exhibit dynamical instabilities. Both energetic and dynamical instabilities have been the subject of intense theoretical and experimental investigations in ultra cold atomic gases~\cite{Wu2001,Smerzi2002,Menotti2003,Fallani2004,DeSarlo2005,Mun2007,Watanabe2009,Wu2006}. 

The recent realization of spin-orbit-coupled gases \cite{Lin2011,Williams2012,Wang2012,Cheuk2012,Zhang2012} is opening new  perspectives in the study of superfluid phenomena. 
These systems lack Galilean invariance and show the consequences also in uniform density configurations. 
In particular, it has recently been shown that in these systems 
the usual Landau criterion for stability, which applies to the motion of an impurity in the medium, cannot be used to determine the stability of configurations carrying a supercurrent, the corresponding critical velocities being dramatically different \cite{Zhu2012,Zheng2012}.

In the present work, we discuss the dynamical as well as the energetic instabilities exhibited by supercurrent states in an ultracold Bose-Einsetin condensed  gas with a type of spin-orbit coupling already realized in experiments~\cite{Lin2011,Zhang2012}. In Sec. II we discuss the connection between the lack of Galilean invariance and the fact that the current is not  conserved in such systems. Then we determine the supercurrent state by adding a Lagrange constraint to the Hamiltonian in Sec. III and explore the instability conditions by calculating the corresponding frequencies of the elementary excitations in Sec. IV. The critical values of the velocity are calculated as a function of the Raman coupling $\Omega$. 
In Sec. V we show  that the collective dipole oscillation, in the presence of a harmonic trap, is well suited for investigation of the effects of instability, and make first comparisons with a recent experiment~\cite{Zhang2012}. Finally, we conclude in Sec. VI.

\section{Hamiltonian and lack of Galilean invariance}

We consider a two-component Bose gas with spin-orbit coupling whose Hamiltonian is given by~\cite{YunLi2012PRL,YunLi2012EPL}
\begin{align}
	&\mathcal{H}
	=
	\mathcal{H}_{\mathrm{int}}+
	\notag \\
	&
	\sum_{\mathbf{p}}
	\begin{pmatrix}
	\psi_{\uparrow, \mathbf{p}}^\dagger & \psi_{\downarrow, \mathbf{p}}^\dagger
	\end{pmatrix}
	\left[
	\frac{(p_x - k_0 \sigma_z)^2 + p_y^2 + p_z^2}{2m} + \frac{\Omega}{2}\sigma_x
	\right]
	\begin{pmatrix}
	\psi_{\uparrow, \mathbf{p}} \\ \psi_{\downarrow, \mathbf{p}}
	\end{pmatrix}, \label{socham}
\end{align}
where $\psi_{\sigma, \mathbf{p}}$ is an annihilation operator of a particle with spin $\sigma$ and momentum $\mathbf{p}$. The Hamiltonian, (\ref{socham}), can be derived 
by applying a local ($x$-dependent) rotation in spin space to the Hamiltonian defined in the laboratory frame, where two detuned spin-polarized laser fields are coupled to the system~\cite{Lin2011,Martone2012}.
The spin-orbit coupling strength $k_0$ is fixed by the momentum transfer of the two lasers, while the Raman coupling $\Omega$ is determined by the laser intensity. The matrices $\sigma_i$, with $i = x, y, z$, are $2 \times 2$ Pauli matrices.
The single-particle part of Hamiltonian (\ref{socham}) has two branches, the lower energy branch exhibiting  two minima if $\Omega < 4k^2_0/2m$.
We assume an $s$-wave interaction with a common intra-species interaction ($g_{\uparrow \uparrow} = g_{\downarrow \downarrow} \equiv g$):
\begin{align}
	\mathcal{H}_{\mathrm{int}}
	=&
	\frac{g}{2V}
	\sum_{\sigma = \uparrow, \downarrow}
	\sum_{\mathbf{p}_1 + \mathbf{p}_2 = \mathbf{p}_3 + \mathbf{p}_4}
	\psi^\dagger_{\sigma, \mathbf{p}_4} \psi^\dagger_{\sigma, \mathbf{p}_3} \psi_{\sigma, \mathbf{p}_2} \psi_{\sigma, \mathbf{p}_1}
	\notag \\
	&+
	\frac{g_{\uparrow \downarrow}}{V}
	\sum_{\mathbf{p}_1 + \mathbf{p}_2 = \mathbf{p}_3 + \mathbf{p}_4}
	\psi^\dagger_{\uparrow, \mathbf{p}_4} \psi^\dagger_{\downarrow, \mathbf{p}_3} \psi_{\downarrow, \mathbf{p}_2} \psi_{\uparrow, \mathbf{p}_1},
\end{align}
where $V$ is the volume of the system.
Despite the presence of the laser fields in the laboratory frame, Hamiltonian (\ref{socham}) in the spin-rotated frame is translationally invariant, the commutation relation $[H,\vec P]=0$ being exactly satisfied, where $\vec{P} \equiv \sum_\mathbf{p} \mathbf{p} \left( \psi_{\uparrow \mathbf{p}}^\dagger \psi_{\uparrow \mathbf{p}} + \psi_{\downarrow \mathbf{p}}^\dagger \psi_{\downarrow \mathbf{p}} \right)$ is the momentum operator. Hamiltonian (\ref{socham})  lacks Galilean invariance~\cite{Zhang2012,Zheng2012}, however. This follows from the fact that the current operator
\begin{align}
	\vec{J}
	\equiv
	\frac{1}{m}\sum_\mathbf{p}
	\begin{pmatrix}
	\psi_{\uparrow, \mathbf{p}}^\dagger & \psi_{\downarrow, \mathbf{p}}^\dagger
	\end{pmatrix}
	\left(
	p_x - k_0 \sigma_z, p_y, p_z
	\right)
	\begin{pmatrix}
	\psi_{\uparrow, \mathbf{p}} \\ \psi_{\downarrow, \mathbf{p}}
	\end{pmatrix}
	\label{currentoperator}
\end{align}
satisfying  the equation of continuity does not coincide with the momentum, due to the presence of the spin term. Furthermore, due to the presence of the Raman coupling in  (\ref{socham}), the current operator does not commute with $H$.

\section{Supercurrent state}
A natural way to construct a supercurrent state is through the determination of the ground state of the Hamiltonian
\begin{align}
	\mathcal{H}_v
	\equiv
	\mathcal{H} - \vec{v} \cdot \vec{P} \label{Hv}
\end{align}
obtained by adding to $\mathcal{H}$ a Lagrange multiplier term proportional to the total momentum $\vec{P}$. Using the operator $\vec{P}$  rather than the current operator $\vec{J}$ in (\ref{Hv}) is crucial in order to ensure that the ground state of $\mathcal{H}_v$  is also an eigenstate of the original Hamiltonian (\ref{socham}). This is guaranteed by the commutativity of $\vec{P}$ with $\mathcal{H}$.

In the present work, we focus on the plane-wave phase, in which only a single momentum state is macroscopically occupied and the density of the system is uniform.
In the plane-wave phase the macroscopic wave function of the condensate takes the form
\begin{align}
	\psi_v
	=
	\sqrt{n}
	\begin{pmatrix}
	\cos \theta \\ -\sin \theta
	\end{pmatrix}
	e^{i\mathbf{k}_1\cdot \mathbf{r}}, \label{psiv}
\end{align}
where $n$ is the density of particles, and $\theta$ and $\mathbf{k}_1$ are variational parameters to be determined by minimizing the energy with respect to $\mathcal{H}_v$.
The angle $\theta$ determines the spin struture of the wave function. In particular, the average spin polarization of the gas is given by $\langle \sigma_z \rangle= \cos 2\theta$, while  $\mathbf{k}_1$ determines the value of the momentum $\vec{P}$, equal to $N\mathbf{k}_1$, with $N$ the number of particles.

\begin{figure}[htbp]
\begin{center}
\includegraphics[width=8.8cm]{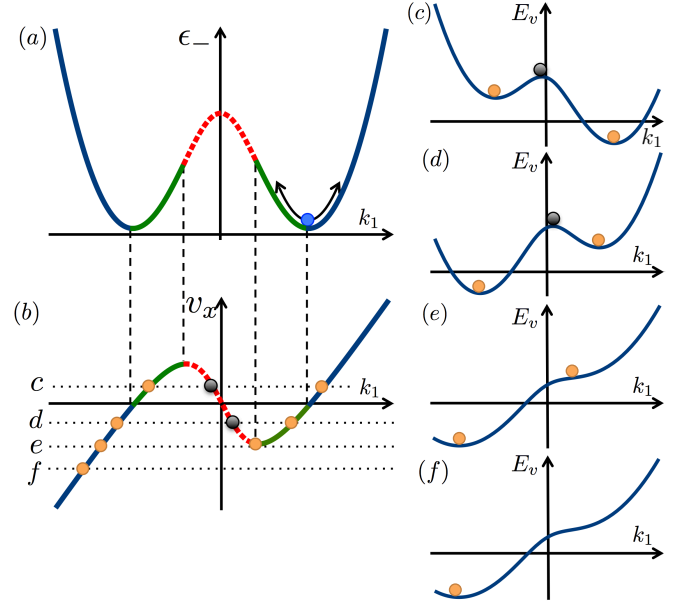}
\caption{(a) Single-particle eigenenergy and (b) corresponding speed as a function of $k_1$ when $\Omega = 1.7 k_0^2/2m$. (c-f) $E_v$  for four values of $v$ [indicated in (b)], as a function of $k_{1}$. Local minima of $E_v$ are represented by orange circles, and local maxima by black circles. (a, b) The dynamically unstable region is shown by the dotted (red) line; the region which corresponds to the metastable states, by green lines; and the lowest energy states of $E_v$, by blue lines. The collective dipole oscillation occurs around one of the minima of the single-particle spectrum, as described in (a).}
\label{envel}
\end{center}
\end{figure}

The energy per particle $E_v=\langle \mathcal{H}_v \rangle/ N$, calculated on the state $\psi_v$, is given by
\begin{align}
	E_v
	=&
	\frac{k_0^2 + \mathbf{k}_1^2}{2m} - \mathbf{k}_1\cdot\vec{v} - \frac{k_{1,x} k_0}{m}\cos 2\theta - \frac{\Omega}{2}\sin 2\theta
	\notag \\
	&\hspace{1cm}+ \frac{gn}{2} - \frac{g-g_{\uparrow \downarrow}}{4}n \frac{\sin^2 2\theta}{4}. \label{ev}
\end{align}

Setting the derivative of $E_v$ with respect to $\mathbf{k}_1$ equal to $0$, one obtains the relationship
\begin{align}
	\mathbf{k}_1 = \left( k_0 \cos 2\theta, 0, 0\right) + m\vec{v}. \label{k1rel}
\end{align}
among the value of $\mathbf{k}_1$, the angle $\theta$, and the velocity $\vec{v}$.
Recalling that $\cos 2\theta$ gives the average polarization one immediately finds that the total current of state $\psi_v$ is given by the expected relationship $\langle \vec{J} \rangle = N \vec{v}$,
revealing that the velocity of the flow in state $\psi_v$ coincides with the velocity  $\vec{v}$  entering the Lagrange constrained Hamiltonian $\mathcal{H}_v$.
In the following, we assume that $\vec{v}$, and hence $\mathbf{k}_1$, is oriented  along the $x$ direction,
and to simplify the notation we set $k_{1,x} \equiv k_1$.

Similarly, setting the derivative of $E_v$ with respect to $\theta$ equal to $0$, one obtains an equation for $\theta$.
This equation   takes a simple form when the interaction is isotropic ($g = g_{\uparrow\downarrow}$), in which case we have
\begin{align}
	\cos 2\theta &= \frac{k_{1}}{\sqrt{k_{1}^2 + (m\Omega / 2k_0)^2}},\label{thetacon}
\end{align}
independent of the value of the interaction coupling constant. In (\ref{thetacon}) 
we have chosen $\sin 2\theta > 0$ and the sign of $\cos 2\theta$ to be the same as that of $k_1$,
which corresponds to choosing the lower branch configuration of the system.
From (\ref{k1rel}) and (\ref{thetacon}), one can express the velocity of the supercurrent state as a function of $k_1$ as
\begin{align}
	v(k_1) = \frac{k_{1}}{m}\left( 1 - \frac{k_0}{\sqrt{k_{1}^2 + (m\Omega / 2k_0)^2}}\right).\label{singlevel}
\end{align}
The stationarity conditions for $k_1$ and $\theta$  give, in general, three solutions [see Figs.~\ref{envel}(c) and~\ref{envel}(d)], one corresponding  to the lowest energy state  (ground state of $\mathcal{H}_v$), where $v$ and $k_1$ have the same sign; a second metastable solution corresponding to a local minimum, where $v$ and $k_1$ have opposite sign; and, finally, a solution corresponding to a local maximum of $E_v$. 
Below, we discuss the stability of these supercurrent configurations.
For this purpose, in the following we focus on the case $g = g_{\uparrow \downarrow}$, which is a good approximation in the available  experiments with rubidium atoms~\cite{Lin2011,Zhang2012}.
It is straightforward to extend the formalism to the case $g \neq g_{\uparrow \downarrow}$.

\section{Instability Conditions}
\subsection{Dynamical instability}

Dynamical instability is characterized by the occurrence of positive imaginary components in the excitation spectrum.
We find the conditions for dynamical instability by looking at the dispersion of the elementary excitations and, in particular, at the sound velocity.
The dispersion of the elementary excitations is obtained by investigating the poles of the Green's function within the Bogoliubov approximation~\cite{Ozawa2012PRL}, taking the condensate wave function,
(\ref{psiv}), with the value of $k_1$ and $\theta$ determined by the solution of (\ref{k1rel}) and (\ref{thetacon}).
(The dispersion can be equivalently obtained using the hydrodynamical formalism~\cite{Martone2012}.)
In the long wavelength (sound) limit we can write the dispersion in the linear form  $\omega \approx c_\pm  |q|$, where $q$ is the momentum shift  relative to $k_1$  and $c_+$
($ c_-$) is the sound velocity when $ q>0$ ($q< 0$). After some straightforward algebra we obtain the result
\begin{align}
	c_\pm = \sqrt{gn \frac{\partial^2 \epsilon_- (k_1)}{\partial k_1^2}} \pm v, \label{soundv}
\end{align}
where 
\begin{align}
	\epsilon_- (k_1)
	=
	\frac{k_1^2 + k_0^2}{2m} - \sqrt{\left( \frac{k_{1} k_0}{m} \right)^2 + \left( \frac{\Omega}{2}\right)^2}
\label{epsilon}
\end{align}
is the lower branch eigen-energy of the single-particle part of Hamiltonian (\ref{socham}).
The sound velocity, (\ref{soundv}), develops an imaginary part when the inverse effective mass $\partial^2 \epsilon_-/ \partial k_1^2$ becomes negative~\footnote{
For pure Rashba spin-orbit coupling, an analogous dynamical instability has been predicted when the momentum of the condensate is lower than the spin-orbit coupling strength~\cite{Zhu2012}.
}. 
The region of dynamical instability is indicated by dotted (red) lines in Figs.~\ref{envel}~(a) and~\ref{envel}~(b).
A similar mechanism of dynamical instability is exhibited by quantum gases in the presence of periodic potentials~\cite{Wu2001,Smerzi2002,Menotti2003,Fallani2004,DeSarlo2005,Mun2007,Watanabe2009,Wu2006}, where the negativity of the effective mass is caused by the band structure of the single-particle
excitation spectrum~\cite{Smerzi2002}.
There is, however, a great difference between the two cases.
In fact, differently from the case of gases in a periodic potential, the density of our systems is uniform and the violation of Galilean invariance is caused by the presence of the spin term in the current operator,~(\ref{currentoperator}).

The spin structure of the order parameter of the stationary solution of $\mathcal{H}_v$ coincides with that of the noninteracting model.
In particular, the velocity, (\ref{singlevel}), of the supercurrent state
is identical to the velocity $\partial \epsilon_- /\partial k_1$ of the single-particle state with momentum $k_1$.
This  correspondence  enables us to understand the  stationary solutions of $E_v$ in terms of the single-particle eigenstates.
As shown in Figs.~\ref{envel}~(a) and~\ref{envel}~(b), there are up to three single-particle eigenstates for a given velocity $v$.
In Figs.~\ref{envel}~(c) to~\ref{envel}~(f), we plot $E_v$, with condition (\ref{thetacon}) for the spin structure of the wave function, as a function of $k_{1}$ for different values of $v$, as indicated in Fig.~\ref{envel}~(b).
When $v$ is positive, the energy minimum state with positive $k_{1}$ is the lowest energy state, while the energy minimum state with negative $k_{1}$ (if it exists) is a metastable state.
One can note that the  local minimum energy states of $\mathcal{H}_v$ correspond to single-particle eigenstates having eigenenergy $\epsilon_-(k_1)$ with a positive curvature.
Conversely, local maximum energy states of $\mathcal{H}_v$ correspond to single-particle eigenstates having eigenenergies with a negative curvature and are dynamically unstable.

\subsection{Energetic instability}

\begin{figure}[htbp]
\begin{center}
\includegraphics[width=8.0cm]{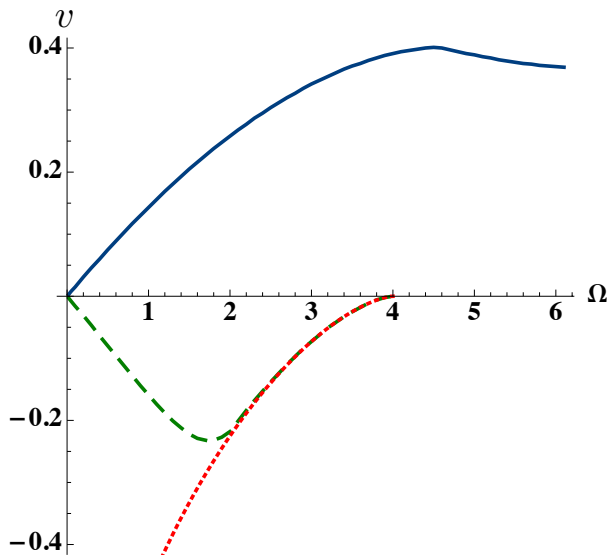}
\caption{Critical velocity for the energetic instability of the supercurrent (in units of $k_0/m$) plotted against $\Omega$ (in units of $k_0^2/2m$) when the condensate has a positive momentum.
The solid (blue) line corresponds to the lowest energy state, and the dashed (green) line to the metastable state, when $gn = 0.48 k_0^2/2m$.
The dotted (red) line represents the onset of dynamical instability.}
\label{criticalv}
\end{center}
\end{figure}

Energetic instability occurs when the system develops a negative excitation energy.
As before, we restrict ourselves to the case where $\vec{v}$ is oriented along the $x$ direction and we consider equilibrium configurations with positive $k_1$.
When $v$ is small, both the lowest energy state and the metastable state of local minimum energy have excitation spectra everywhere positive. However, above a certain critical  value of $v$, the excitation spectrum starts taking negative values. 
Figure~\ref{criticalv} plots the critical velocity as a function of $\Omega$.
For the interaction parameter, we take $gn = 0.48 k_0^2/2m$, which is the value used in~\cite{Martone2012} and is relevant in the presently available experiments~\cite{Zhang2012}.
The solid (blue) line is the critical velocity calculated assuming that the system is in the minimum energy state ($v$ and  $k_1$ have the same sign), which was calculated in~\cite{Zheng2012}. The dashed (green) line instead corresponds to the metastable configuration ($v$ and $k_1$ have the opposite sign).
The metastable state does not exist below the dotted (red) line, corresponding to the onset of the dynamical instability, which approaches the value $-k_0/m$ as $\Omega \to 0$.

We find that, when $\Omega$ is small, the instability of the lowest energy state and the metastable state is due to the roton part of the excitation spectrum.
At large $\Omega$, the instability is instead caused by the phonon branch.
The transition between roton instability and phonon instability for the lowest energy state occurs around $\Omega \approx 4.6 k_0^2/2m$, where one can see a kink in the figure.
In the absence of the supercurrent, the phase transition between the plane-wave phase with nonzero momentum and the condensate with zero momentum takes place at $\Omega = 4.0 k_0^2/2m$,
and when the condensate occupies the zero-momentum single-particle state, the excitation spectrum does not exhibit any roton structure.
Nevertheless, in the interval $4.0 k_0^2/2m < \Omega < 4.6 k_0^2/2m$ instability of the supercurrent occurs due to the emergence of roton excitations, which is characteristic of the plane-wave phase. 
The transition between roton and phonon instabilities for the metastable state occurs at smaller values of the Raman coupling, around $\Omega \approx 2.0 k_0^2/2m$.
The maximum (in magnitude) velocity compatible with the existence of a metastable state [dotted (red) line in Fig.~\ref{criticalv}; also see Fig.~\ref{envel} (e)] approaches 0 as $\Omega$ reaches $\Omega = 4.0 k_0^2/2m$ from below; above $\Omega = 4.0 k_0^2/2m$ there is no metastable state.
If we introduce a small difference in the interaction parameters $g \neq g_{\uparrow \downarrow}$, 
the overall structure does not change in a significant way, except at small values of $\Omega$, where the critical  velocities vanish  at a finite value of $\Omega$, corresponding to the transition  to the  stripe phase~\cite{YunLi2012PRL}.

\begin{figure}[htbp]
\begin{center}
\includegraphics[width=8.0cm]{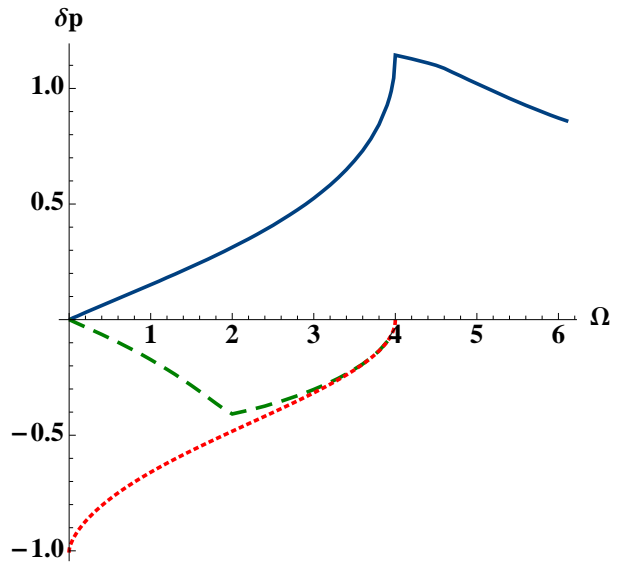}
\caption{Critical momentum kick $\delta p$ to the condensate which causes instabilities (in units of $k_0$) as a function of $\Omega$ (in units of $k_0^2/2m$). The solid (blue) line and the dashed (green) line show the critical momentum required to reach energetic instability for the lowest energy state and the metastable state, respectively, when $gn = 0.48 k_0^2/2m$. The dotted (red) line represents the critical momentum kick required to get dynamical instability.}
\label{kickvomega}
\end{center}
\end{figure}

One can also discuss the instability in terms of the momentum kick (or momentum excess) that one can provide to  the condensate, initially at equilibrium~\footnote{
Consequences of the lack of Galilean invariance in spin-orbit-coupled Fermi gases can also be investigated through the momentum kick, as discussed in~\cite{Maldonado-Mundo2013}.
}. 
In Fig.~\ref{kickvomega}, we plot the minimum momentum kick $\delta p = p - k_{\mathrm{eq}}$ required to reach the unstable region as a function of $\Omega$.
Here, $k_\mathrm{eq}$ is the equilibrium value of the momentum $k_1$ [$k_\mathrm{eq} = k_0\sqrt{1-(m\Omega/2k_0^2)^2}$ for $\Omega \le 4.0 k_0^2/2m$ and $k_\mathrm{eq} = 0$ for $\Omega > 4.0 k_0^2/2m$].
The solid (blue) and dashed (green) lines correspond to the value of $\delta p$ required to reach the energetically unstable region for positive and negative kicks, respectively. 
The dotted (red) line in Fig.~\ref{kickvomega} shows the critical momentum kick needed to reach the dynamically unstable region.

Let us, finally, notice that, as anticipated in Sec. I, the critical velocity of the supercurrent differs from the value predicted using the Landau criterion because of the breaking of Galilean invariance ~\cite{Zheng2012}. Differently from the supercurrent instabilities, which can exhibit either energetic or dynamical instability, Landau instability always has an energetic nature. 

\section{Dipole oscillation.}

We expect that the instability discussed above can be observed experimentally through the excitation of the collective dipole oscillation in a harmonic trap.
Actually in our systems the center-of-mass motion is coupled to the spin degrees of freedom by the Raman coupling, and Kohn's theorem is no longer applicable~\cite{YunLi2012EPL}.
The dipole mode corresponds to an oscillation of the condensate around one of the minima of the single-particle spectrum [Fig.~\ref{envel}(a)], and during the oscillation the condensate behaves as the supercurrent state discussed above with alternating velocity.
As one can see from Fig.~\ref{kickvomega}, if we provide a sufficiently large momentum kick to the condensate, the system will enter a dynamically unstable region and the oscillation will break down.
Indeed in~\cite{Zhang2012}, where dipole oscillation was excited by suddenly transferring a momentum kick to the condensate, the authors could not observe a single-frequency oscillation in the region $2.5 k_0^2/2m< \Omega < 4.0 k_0^2/2m$.
Our estimate from Fig.~\ref{kickvomega} shows that if the initial momentum kick is $|\delta p| \approx 0.4 k_0$, the condensate actually becomes dynamically unstable in the  above region.

Observing energetic instability is, in general, more challenging than
observing dynamical instability because thermalization and collisional
effects are required to trigger it. Furthermore the corresponding critical
velocity depends on the value of the density and becomes lower and
lower as one approaches the border of the condensate.

\section{Conclusion}

In conclusion we have shown that a remarkable consequence of spin-orbit
coupling is the occurrence of dynamical instability in the superfluid flow
when its velocity reaches a critical value. Contrary to the usual
superfluids, this unexpected feature also occurs for configurations
with a uniform density and is the consequence of the violation of Galilean
invariance caused by the spin-orbit coupling.

\begin{acknowledgments}
We thank Shuai Chen and Hui Zhai for clarifying discussions of their experiment~\cite{Zhang2012}.
T. O. thanks Zeng-Qiang Yu for useful discussions of Ref.~\cite{Zheng2012}.
This work was supported by the ERC through the QGBE grant and by Provincia Autonoma di Trento.
\end{acknowledgments}

\end{document}